\documentclass[12pt]{article}

\usepackage{graphicx,amsmath}

\title{Theory of the deconfinement in QCD\thanks{v2 --  accepted for publication in Phys. Atom. Nuclei,\ 29-09-23}}
\author{M. S. Lukashov\thanks{m.s.lukashov@gmail.com} and Yu. A. Simonov\thanks{simonov@itep.ru}  \\
NRC ``Kurchatov Institute'' -- KCTEP \\
Moscow, Russia}

\newcommand{\beq}{\begin{eqnarray}}
 \newcommand{\eeq}{\end{eqnarray}}
\newcommand{\be}{\begin{equation}}
 \newcommand{\ee}{\end{equation}}

\def\fun#1#2{\lower3.6pt\vbox{\baselineskip0pt\lineskip.9pt
\ialign{$\mathsurround=0pt#1\hfil ##\hfil$\crcr#2\crcr\sim\crcr}}}

\newcommand{{\SD}}{\rm SD}

\newcommand{{\Mc}}{\mathcal{M}}

\newcommand{\lan}{\langle}
\newcommand{\ran}{\rangle}

\begin{document}
\maketitle
\begin{abstract}
The phenomenon of the deconfinement -- the spectacular drop  of the colorelectric string tension at the critical temperature $T_c$-is studied within
the method of field correlators (FCM)  taking into account directly the contribution of the gluon condensate into the hadronic free energy. Using the resulting expressions for the free energy as a sum of the gluon condensate (the vacuum energy) and the hadronic pressure one obtains the possibility to calculate the
deconfinement temperature $T_c$ and the temperature behavior of the string tension $\sigma_E(T)$ and the gluonic condensate $G_2(T)$ below $T_c$. The
connection between the string tension and the quark condensate found in the framework of FCM allows the predict also the latter as a function of $T$ . These results are  compared to the known lattice data of $T_c$, $\sigma_E(T)$, $\langle \bar{q}q \rangle(T)$  for hadronic media with different $n_f$ and $m_q$ and in the external magnetic field $eB$. The good agreement of the results of this approach  with lattice data is demonstrated.
\end{abstract}

\section{Introduction}

The phenomena of the confinement and deconfinement in QCD are the basic properties of our nature and define to the great extent
the physics in the Universe. It is known that confinement yields more than $90$ percent of the visible mass in the universe
and it disappears at very large temperatures which can be seen in the emission behavior of large stars.
On the theoretical side the theory of confinement is in the process of establishing: the most developed theory based on the
basis of  field correlators-the method of field correlators (FCM) \cite{1,2,3,4,5} explains all confinement phenomena in terms of vacuum averages of bilocal colorelectric field strength which are calculated selfconsistently via themselves so that the only parameter of confinement- the string tension $\sigma$ defines all confinement events. This approach was checked in comparison with lattice data \cite{6,7,8,9,10} and is fully supported in all details. It is important that string tension in FCM is expressed mostly via squared field strength $F_{\mu\nu}$ (higher powers of $F$ yield a few percent of string tension \cite{11}) and therefore
one obtains the so-called Casimir scaling for higher $SU(3)$ representations of gluonic matter \cite{11} which is in a good agreement with lattice data \cite{12,13,14}.
It is tempting to try to find a more elementary source of confinement -- not bilocal field strength correlator, but some "elementary" fields or objects (like magnetic monopoles, instantons, etc.) -- these attempts are continuing for the last decades -- see \cite{15,16,17,18} for some basic papers -but have not yet brought a well defined theory in a good agreement with lattice and experiment -- e.g. the center vortex model strongly disagrees with lattice data in particular in the Casimir scaling prediction \cite{18*} and in the resulting string tension value \cite{18**}.
The problems of deconfinement and of the deconfined phase of QCD have been the topic of intense  numerical and theoretical studies \cite{19,20,21,22,23,24,25,26,27,28,28*,29,30,31,32,33,34}. In the framework of the FCM the problem of the temperature phase transition  and the QGP dynamics was
studied in \cite{35,36,37,38,39} and the problem of the colormagnetic confinement (CMC) was reviewed recently in \cite{40}.

While phenomena of the colorelectric  and colormagnetic confinement were quantitatively explained and exploited within the FCM \cite{1,2,3,4,5,6,7,8,9,10,40} the phenomenon of the deconfinement -- the sharp decrease and vanishing of the colorelectric string tension $\sigma(T)$ near $T_c$ found on the lattice \cite{30,34} -- is not yet explained analytically. Nevertheless
 the basic deconfinement problem -- the deconfinement temperature $T_c$ -- was  quantitatively calculated within the formalism where
  the total free energy contains the vacuum energy term -- the gluonic condensate \cite{35,40*,41,42}. The resulting values of $T_c$  were found  in a good agreement with the lattice data \cite{43,44,45} also with account of different quark masses,magnetic field and baryon density $\mu$.
It is our aim below to extend this formalism and  define the behavior of the gluonic condensate and the string tension as functions of $T$ also below $T_c$  taking into account  pressure, energy density,quark and hadron masses,external fields -- all thermodynamic characteristics of the hadronic and quark-gluonic media in this region. This will be the main purpose of this paper.

In all cases the type of deconfinement transition is defined by the combination of the basic properties of the string tension
obtained from the field correlators \cite{46,47,48} -- and by the thermodynamic ensemble of the QCD matter. In what follows we shall discuss both fundamental microscopic and thermodynamic aspects of the deconfinement phenomenon. Here the basic role is played
by the vacuum energy density -- gluon condensate $\epsilon$ which is a fundamental concept in the dynamics of elementary particles as was suggested in \cite{49,50,51,52}.
The next important step in this direction was done in \cite{53} where the colorelectric string tension $\sigma_E(T)$ was found to be connected with the gluonic condensate $\epsilon$ and in this way all confined dynamics is directly defined by the gluonic condensate.
In our formalism below (as well as in the previous studies \cite{35,40*,41,42})
the gluon condensate $\epsilon$ directly enters in the free energy of the confined matter
via the colorelectric vacuum energy density $1/2 \epsilon(T)$ namely,
 as was
suggested and studied in \cite{41,53}, the free energy in the confined phase $F_1(T)$ is defined as
\be
|F_1(T)|=  |\epsilon_{vac}(T)| + P_h(T), \epsilon_{vac}= 1/2 \epsilon_g + \epsilon_q, \epsilon_q= \sum_q m_q \langle \bar q q \rangle
\label{01} \ee
where the gluon condensate enters as follows
\be
\epsilon_g(T)= -\frac{b \alpha_s \langle G^2 \rangle}{32\pi},\langle G^2 \rangle=\langle G^a_{\mu\nu}G^a_{\mu\nu} \rangle . \label{02} \ee
We have taken into account in the eq.(\ref{01}) that only the colorelectric gluon condensate $\langle G_E^2 \rangle= 1/2 \langle G^2 \rangle$ is connected with the pressure which is accounted for by the coefficient $1/2$ before $\epsilon_g(T)$. We also note that the quark condensate
term $\epsilon_q$ as found in \cite{41} contributes around $(10-15)$ percent of the total vacuum energy and disappears at the same temperature $T_c$ and therefore below we disregard it in the first approximation.
In eq.(\ref{01}) $P_h(T)$ is the hadron interacting gas pressure growing with the temperature while the gluon condensate $\langle G^2(T) \rangle$ (its colorelectric part) decreases, vanishing at $T=T_c$. At the same time the colormagnetic part develops independently and finally grows at large T. In this way the colorelectric and colormagnetic d.o.f. are disconnected (in the first approximation).

At this point one must define the behavior of the vacuum energy $\epsilon_g(T)$ as a function of the temperature which will
explain the properties of the hadron gas and its deconfinement transition. In what follows we impose the following condition
on the confining free energy which will be called The Vacuum Dominance Mechanism (VDM) where the hadronic pressure is growing with
temperature $T$ with the simultaneous decrease of the vacuum energy (gluon condensate), so that their sum is kept constant.
\be
|F_1(T)|= 1/2 |\epsilon(T)| + P_h(T)= 1/2|\epsilon(T=0)|. \label{03} \ee
In this way the basic QCD quantity -the gluon condensate \cite{49} -- $\langle G_E^2(0) \rangle= \frac{32\pi \epsilon(0)}{b\alpha_s}$  defines the properties of the deconfining process. In particular, taking eq.(\ref{03}) at $T=T_c$  and accounting for the equality
$P_h(T_c)= P_{qgl}(T_c)$ one obtains the equation which defines
the deconfining temperature $T_c$

\be
P_q(T_c) + P_g(T_c)= 1/2|\epsilon_g(0)|.
\label{04} \ee

Using the numerical value of the $|\epsilon_g(0)|$ from \cite{49} and the functions $P_q(T_c),P_g(T_c)$ one can find the values
of $T_c$ for all types of hadronic gas. This latter condition (without the VDR connection eq.(\ref{03}) was used in \cite{35,40*,41,42} to predict the values of $T_c$  and
its dependence on $m_q,\mu,eB$ in good agreement with lattice data. In what follows we shall exploit the relation (\ref{03}),(\ref{04}) to predict
also the behavior of $\epsilon_g(T),\sigma(T)$ and compare it with data.

The main purpose of this paper is to answer the following questions:
\begin{description}
\item
{(1)} What is the deconfinement temperature $T_c$ of the vanishing string tension for different thermodynamic systems.
\item
{(2)} How the colorelectric (CE) string tension decreases with temperature and moreover why it decreases so fast before vanishing at $T=T_c$?
\item
{(3)} Why the CE string tension depends strongly on the contents of the thermodynamic ensemble, so that $T_c(n_f=0) \approx 1.5 T_c(n_f=2)$?
\item
{(4)} What is the connection between the chiral condensate and string tension VS lattice data.
\end{description}

In what follows we shall find the answers to these questions within our approach .
 The plan of the paper is as follows:
 In section 2 the basic role of the vacuum average of the gluonic condensate $G_2$ in the QCD thermodynamics is discussed and the fundamental connection between $G_E^2$ and the CE string tension $\sigma_E(T)$ found earlier is exploited to find the temperature dependence of the latter.
 In section 3 the deconfining temperature $T_c$ is defined using  eq.(\ref{04}) for different systems using both new and old results for  hadronic systems with different quark masses and also in magnetic field and nonzero chemical potential $\mu$.
In section 4 we use the string tension  connection to the basic characteristics of the QCD
vacuum-the gluon condensate was found in \cite{53} and discussed in section 2 where the gluon condensate is proportional to the square of the string tension  (the details of this connection are given in the appendix 1). As a result the sharp T-dependence of the string tension $\sigma_E(T)$ is found and a good agreement with the lattice data is demonstrated in Fig. (\ref{fig04}) in this section.
In section 5 the temperature behavior of the quark condensate below $T_c$ is defined using its connection with the string tension and compared with lattice data.
The concluding section contains a summary of results and a discussion of possible developments of the deconfinement theory.

\section{The string tension vs gluonic condensate}

The internal structure of confinement is defined by the so-called gluelump Green's functions $G^{(2gl)}(z)$ which are colorelectric propagators of two gluons  on distance $z(x,y)$ with the fixed Wilson line between $x,y$ to make the whole system gauge invariant \cite{46,47,48}. The colorelectric field correlator $D^{E}(z)= \frac{g^4(N^2_c -1)}{2} G^{(2gl)}(z)$ defines the colorelectric (CE)
string tension $\sigma^E (T) \equiv \sigma(T)$ which will be the main object of discussion below,
\be
\sigma^{E}(T)= 1/2 \int d^2 z D^{E}(z). \label{04*} \ee
At this point it is important to stress the difference between colorelectric (CE) and colormagnetic (CM) field correlators $D^{E}(z)$ and $D^{H}(z)$, gluon condensates $G^E_2= \frac{\alpha_s}{4\pi} \langle (G^a_{i4})^2 \rangle$,and $ G^H_2= \frac{\alpha_s}{4\pi} \langle (G^a_{ik})^2 \rangle$ and string tensions $\sigma^E(T), \sigma^H(T)$. As can be seen in explicit expressions and  was observed in lattice calculations \cite{6,7,8,9,10} the two types of phenomena -- CE and CM -- and their magnitudes are only weakly connected and the dynamics in both has
completely different character: e.g. $\sigma^H(T)$ is almost constant in the interval $0<T<T_c$ (where $\sigma^E(T)$ vanishes)
and grows quadratically beyond this interval. Therefore we shall neglect in what follows the influence of the CM components and
consider below only the CE functions, omitting the index CE for simplicity.
It is important that the interaction kernel in $D^{E}(z) \equiv D(z)$ is again due to confinement strings between gluons and the Wilson line and therefore for large distances as shown in \cite{53} one has a check of selfconsistency where the string tension is expressed
via string tension of the internal strings so that these factors cancel on both sides of the eq.(\ref{02}) for distances $r > 1$ GeV$^{-1}$ . However as shown in  \cite{53} the basic connection arises at small distances $r < 1$ GeV$^{-1}$ where string tension can be expressed via the vacuum average of the gluonic condensate.
 Indeed $D(z)$ has a humpback structure as shown in Fig. 1 of \cite{53} with a maximum at $z= z_0 =1$ GeV$^{-1}$ and studying its structure in \cite{53} one obtains
\be
D(0)= 0.15 D(z_0)= \frac{\pi^2}{2} G_2 = 0.156 4\pi^2 \alpha^2_v \sigma^2(T) \exp{(-M_0 z_0)} \label{05} \ee
where $M_0=2.4$ $GeV$ is the lowest gluelump mass found both in theory \cite{35,36} and on the lattice \cite{39,40}.
As a result one obtains the important relation between the string tension and the gluonic condensate $G_2(T)$
\be
G_2(T)= \alpha_s/\pi \langle 0|(G^a_{\mu\nu})^2|0 \rangle= \alpha_s/\pi \langle G^2 \rangle = 1.69 \sigma^2 \alpha_s^2. \label{06} \ee
At this point one takes into account that the colorelectric (CE) $\sigma$ can be connected only to the CE part of $\langle G^2 \rangle$
\be
\sigma_E(T)= \sqrt{\frac{1/2 G_2(T)}{1.69\alpha_s^2}} \label{07} \ee
and normalizing at $T=0$  one obtains $\sigma^2(T) \approx 5.4 G_2(T)$
where we neglect the $T$ dependence of the $\alpha(T)$.

In this way we have connected the gluonic condensate $G_2(T)$ with the string tension $\sigma(T)$ and in what follows we shall
be able to deduce the decrease of $\sigma(T)$ for $T \approx T_c$ using  the corresponding decrease of the $G_2(T)$ connected to the basic EoS of the hadron medium.

Now omitting first the quark mass terms one can see that in the case when the effective pressure $\frac{P_h(T)}{T^4}$ is growing with $T$  the gluon condensate will strongly decrease with $T$, which means gradual deconfinement.
In this approach  the gluonic condensate enters directly in the expression for the pressure in the confined phase as was
suggested and studied in \cite{51,52,53} and shown in (\ref{01}) and (\ref{02}).
We have taken into account in the eq.(\ref{06}) that only the colorelectric gluon condensate $\langle G_E^2 \rangle$ is connected with the pressure which is accounted for by the coefficient $1/2$ before $\epsilon$.
Using (\ref{03}) and (\ref{04}) one obtains the following connection of the gluon condensate $\langle G^2 \rangle (T)$ with the pressure $P_h(T)$

\be
\langle G_E^2 \rangle(T)= \langle G_E^2 \rangle(0) - \frac{64 \pi}{b \alpha_s} P_h(T).
\label{08} \ee

This equation allows to find the T dependence of the gluon condensate up to its vanishing at $T=T_c$. Now due to eq.(\ref{06})
one can find the fast decreasing behavior of $\sigma_E(T)$  in the same region

\be
\sigma^2_E(T)= \sigma^2_E(0) - \frac{\xi 64 }{b} P_h(T), \xi \approx 5.4. \label{09} \ee

The corresponding behavior  for different $n_f$  in comparison with lattice data will be demonstrated in section 4.

At this point the main problem is the hadron pressure and its dependence on temperature $T$ and string tension $\sigma(T)$.
We start with the simplest case of  hadrons with mass $m_i(T)$ depending on $T$ due to the confining string tension $\sigma(T)$.
One can write for the noninteracting system of mesons or glueballs the Hadron Resonance Gas pressure
\be
P_h(T)= \sum_i P^i_h(T)= \frac{g_i T^2}{2\pi^2} \sum_{n=1,2,..}\frac{m_i^2 K_2(nm_i/T)}{n^2} \label{10} \ee
where $g_i$ is a multiplicity of hadrons of the type $i$ , $m_i= m_i(T)$ is the hadron mass depending on $T$ via the string tension $\sigma(T)$ and $K_2(z)$ is  $K_n(z),n=2$ the Kelvin special function.
Here one can use the following representation with the definition $z= m/T$
\be
\sum_{n=1,2,...}\frac{K_2(z)}{n^2}= \frac{1}{3z^2}\int_0^{\infty}\,dt\,\frac{t^4}{f(t,z)(\exp((f(t,z)))-1)},
f(t,z)=\sqrt{t^2+z^2} \label{11} \ee
and as a result one obtains
\be
P_h(T)=\sum_{i}\frac{g_i T^4}{6\pi^2}\Phi(z_i),\quad \Phi(z)=\int_0^{\infty}\,dt\,\frac{t^4}{f(t,z)(\exp{(f(t,z))}-1)}.\label{12} \ee

We now turn to the string tension behavior, where the normalized string tension is connected to $G_2(T)$ as in eq.(\ref{05}) $\sigma(T)^2= 5.4 G_2(T)$ and using the relation $\epsilon(T)= b/32 G_2(T)= \frac{b \sigma^2(T)}{172.8}$

one obtains the connection between the decreasing string tension $\sigma(T)$ and the growing hadronic pressure $P_h(T)$
\be
\sigma^2(T)= c_B (P_2(T_c)- P_h(T)), c_B=345/b . \label{13} \ee
The (\ref{12}) and (\ref{13}) show the deconfining process with the growing $T$ in the approach EoS and will be used below to demonstrate this process in comparison with lattice data.

\section{The deconfining temperature $T_c$ for different hadronic systems }

As one can see one can use in the latter case the pressure $P_2(T)$ of the deconfined quark-gluon plasma (qgp) which is easier
to define theoretically as compared to the complicated hadronic matter with the T-dependent masses due to $\sigma(T)$.
The same can be  true in our approach  at least for $n_f>0$ where one can expect the equality of the pressure derivatives
at the point $T=T_c$. To this end we define the pressure of the qgp  following the nonperturbative thermodynamics theory in \cite{41}
\be
P_q= \frac{2N_c T^2 m_q^2}{\pi^2} \sum_1^{\infty}\frac{(-)^{n+1}L^n}{n^2}K_2(nm_q/T) \label{14} \ee
where $L=\exp{(-V_1(\infty)/2)}$,while the gluon gas pressure is
\be
P_g= \frac{2(N_c^2 -1)}{\pi^2} \sum_1^{\infty}\frac{L^n_{adj}T^4}{n^4} \label{15} \ee
and $L_{adj} = L^{9/4}$.

It is also convenient to exploit the simple expressions for the pressure of the massless noninteracting quarks and gluons
\be
P^{(0)}_q= (7/180)\pi^2 N_c n_f T^4, P^(0)_g= (\pi^2/45)(N_c^2 -1)T^4 . \label{16} \ee
Using these equations and (\ref{12}),(\ref{13}) one can calculate deconfining temperatures $T_c$ for all types of qgp,as will be shown. In what follows we are using eq.(\ref{07}) in the following form
\be
1/2\epsilon(T=0)= \frac{b\alpha_s \langle G^2(0) \rangle}{64\pi}=\frac{b G_2(0)}{64}= P_q(T_c) + P_g(T_c). \label{17} \ee
We start with the simplest case of heavy quarks and antiquarks of one flavor with mass $m_q$ ,where eq.(\ref{14}) yields

\be
P_q(T_c)=\frac{2N_c T^2 (m_q)^2L(T)K_2(m_q/T_c)}{\pi^2}=1/2\epsilon(T=0). \label{18} \ee

Using asymptotic form of the $K_2(z)$ and the estimate of $L(T) \approx \exp(-0.3 GeV/T)$  one obtains for $T_c$

\be
T_c= \frac{m_q + 0.3 GeV}{ln(753.6) T_c^2 \sqrt{T_c/m_q})}. \label{19} \ee

As a result for the $m_q=1.4 GeV$ one obtains $T_c= 0.40 GeV$  and for $m_q= 1 GeV$ one gets $T_c \approx 0.3 GeV$.
For a detailed study of the $T_c(m_q)$ dependence in the framework of the present theory see \cite{41} and in particular  Table 2 therein.
A similar picture one can see in the dependence of $T_c$ on the $ mass(PS) \approx 2m_q$ in \cite{34}.

We  consider now the hadron-qgp transition with $n_f>0$, when one expects the continuos pressure and its
derivative at $T=T_c$. We consider first the simplest case of massless quarks when the total QGP pressure is equal to the sum
$P_2(T)= P^{(0)}_q(T) + P^{(0)}_g(T)$ where $P_q,P_g$ are from eqs.(\ref{13}),(\ref{14}). We obtain

\be
P_2(T)= P^{(0)}_q(T) + P^{(0)}_g(T) = C_0 T^4, C_0= (7/180)\pi^2 N_c n_f + (\pi^2/45)(N_c^2 -1). \label{20} \ee

It is interesting that our approach in this case yields quite reasonable results. Indeed, with the same $P_2(T)$ from (\ref{20}) one obtains

\be
T_c^4= \frac{\eta  b G_2(0)/32}{C_0} \label{21} \ee

where $\eta=1/2$ for the colorelectric part of the condensate $G_2= \alpha_s/\pi \langle G_{\mu\nu}G_{\mu\nu} \rangle$ which according to
\cite{49} can be equal to $0.012$ GeV$^4$ for QGP and 3-4 times larger for pure gluon plasma. As a result one obtains
$T_c= 0.15$ GeV for $n_f= 2$ and $T_c= 0.134$ GeV for $n_f=4$ and $T_c= 0.24$ GeV for $n_f=0$ and 3 times larger gluon condensate. These results are in the remarkable agreement with the lattice data \cite{54,55,56,57,58} which yield $T_c(n_f=0) \approx 0.26-0.27$ GeV and $T_c(n_f=2)= (0.16-0.17)$ GeV.
One can continue this analysis for the QGP with nonzero quark masses and Polyakov line interaction as in eqs.(\ref{14}),(\ref{15}). In our approach  one again obtains an interesting
agreement with the lattice data \cite{54,55,56,57,58} which is clearly seen in the Table 1 from \cite{41},which we displace below

\begin{table}
\caption{Transition temperature $T_c$ for massless quarks, $n_f=0,2,3$ (the upper part), and for different nonzero $m_q$ and $n_f$ (the lower part) in comparison with lattice data.}
\vspace{1cm}
\begin{center}
\label{tab01}
\begin{tabular}{|c|c|c|c|c|c|}
  \hline
  $n_f$ & $m_u,$ MeV & $m_d,$ MeV & $m_s,$ MeV & $T_c,$ MeV & $T_c$ (lat), MeV \\
  \hline
  0 & - & - & - & 268 & 276 \cite{54}  \\
  2 & 0 & 0 & - & 188 &  \\
  3 & 0 & 0 & 0 & 174 &  \\
  \hline
  2 & 3 & 5 & - & 189 & 195-213 \cite{55} \\
  2+1 & 3 & 5 & 100 & 182 & 175 \cite{56,57} \\
  3 & 100 & 100 & 100 & 195 & 205 \cite{58}\\
  \hline
\end{tabular}
\end{center}
\end{table}

\begin{table}
\caption{Quark mass dependence of transition temperature with $|\lan \bar{q}q \ran| = (0.13$ GeV$)^3$ in comparison with the lattice data from \cite{54}.}
\vspace{1cm}
\begin{center}
\label{tab02}
\begin{tabular}{|c|c|c|c|c|c|c|c|}
  \hline
  $m_q$, MeV & 25 & 50 & 100 & 200 & 400 & 600 & 1000 \\
  \hline
  $T_c$ (lat), MeV & 180 & 192 & 199 & 213 & 243 & 252 & 270 \\
  $T_c$, MeV & 179 & 185 & 195 & 213 & 245 & 273 & 320 \\
  \hline
\end{tabular}
\end{center}
\end{table}

Here in  the upper part of the Table 1 for the values of $n_f>0$  the eq.(\ref{21}) was used, while in the lower part of the
Table 1 the additional contribution of the quark vacuum energy $\epsilon^q_{vac}= \sum^{n_f}_{i=1} m^q_i | \langle \bar q_i q_i \rangle |$ was used  in \cite{41} as prescribed in \ref{01}. One can see a reasonable agreement of our results with the lattice data for all quark masses.
An additional check of the validity of this theory is given in the \cite{41} for the dependence of $T_c{m_q}$. As it was shown there the combined vacuum energy $\epsilon^q_{vac}$ containing the quark mass produces a
significant shift of the critical temperature, namely

\be
T_c^{(q)}= T_c^{(0)} (1 + \frac{m_q^2}{16(T_c^{(q)})^2}).  \label{22} \ee

The resulting strong dependence of $T_c$ on $m_q$ is shown below in the Table 2 from \cite{41} in a good agreement with lattice data of \cite{33}.

 It is of a great interest also to find the critical temperature for the nonzero baryon density $\mu$. This topic was studied
 in \cite{60} where also nonzero chemical potential was considered. These results can be seen in Fig. 1 and Fig. 2 respectively in \cite{60}.

In this way we come to the conclusion that our approach  allows to obtain the reasonable values of $T_c$ for different values
of the quark masses, number of flavors, magnetic field and chemical potential for different hadron media and in particular for the pure gluonic plasma.

\section{The temperature behavior of the string tension and the gluonic condensate below $T_c$}

We shall discuss here the temperature behavior of $\sigma(T)$, $G_2(T)$ below $T_c$ using eqs.(\ref{08}),(\ref{12}),(\ref{13}) for our approach. At this point it is important to stress that the VDM relations involve the hadron pressure $P_h(T)$ (and the quark-gluon pressure in equations for $T_c$), which can be taken either from theory or from the lattice or experimental data. In what follows and above for the calculation of $T_c$ we have used the theoretical expressions for $P_h(T)$, $P_{qg}(T)$ and have found a good agreement of $T_c$ with the lattice data. However these theoretical pressure expressions do not take into account many physical factors, e.g. the decreasing of hadron masses $M_h(T)$ in the pressure due to the gradual vanishing of the string tension $\sigma(T)$ near $T_c$ ,the similar but much smaller change of $M_h(T)$ due to the color Coulomb correction with $\alpha_s(T)= 2\pi/(b \log(T/\Lambda))$. The analysis shows that the expressions obtained below for the string tension and the color condensate are only the first approximations and the account of decreasing $M_h(T)$ yields a fine structure correction in $\sigma(T)$, $\langle \bar{q}q \rangle (T)$ nearby $T=T_c$ which will be discussed in future publications. Below we shall exploit the theoretical expressions for the pressure of the gas of noniteracting hadrons and non-interacting quarks and gluons for $T$ below or equal to $T_c$ where it describes lattice data with a reasonable accuracy for the combination
 $\eta(T)= \frac{P_h(T)}{P_h(T_c)}$ needed for the determination of $\sigma(T)$, $\langle \bar{q}q \rangle (T)$.

Having defined the values of $T_c$ and $P_h(T)$ one can use the growth of the hadronic pressure $P_h(T)$ from
eqs.(\ref{10}),(\ref{11}),(\ref{12}) for the temperature behavior of $\sigma(T)$ exploiting eq.(\ref{08}).
We start with the case of  hadrons, e.g. $J/\psi$ mesons to compare with the lattice data of \cite{30} and take the hadron pressure in the form of eq.(\ref{10}) neglecting for large $M/T$ higher terms with $n>1$ and approximating $K_2(z) \approx \sqrt{\pi/2z} \exp{-z}$

\be
P_h(T)= \frac{gT^2M^2 K_2(M/T)}{2\pi^2} \approx \frac{g T^{5/2}M^{3/2}}{(2\pi)^{3/2}}\exp(-M/T). \label{23} \ee

Another form of the hadron pressure is given by eq.(\ref{11})
$P_h(T)= \frac{gT^4\Phi(z)}{6\pi^2}$, where $\Phi(z=M/T)$ is given in eq.(\ref{12}). As a result using eq.(\ref{13}) one obtains for the whole region in the interval $[0,T_c]$  from eq.(\ref{08}) that the behavior of $\sigma_E(T)$ is

\be
\sigma^2_E(T)= \sigma^2_E(0) ( 1 - P_h(T)/P_h(T_c)). \label{24} \ee

One can see in eq.(\ref{12}) that $P_h(T)=T^4 Z(M/T)$ where $Z(M/T)$ is finite for $M=0$ and therefore one can write
 for $T$ near $T_c$ neglecting the $Z'(T_c)$ contribution

\be
\sigma^2_E(T)=\sigma^2_E(0)(1-(T/T_c)^4). \label{25} \ee

One can see that the fast growth of the $P_h(T)$ near $T_c$ in eq.(\ref{25}) makes the transition curve of $\sigma(T)$ very steep for both light and heavy  hadrons which can be also seen in lattice data of \cite{30} in Fig. 1.


\vspace{1cm}

\begin{figure}[!htb]
\begin{center}
\includegraphics[width=80mm,keepaspectratio=true]{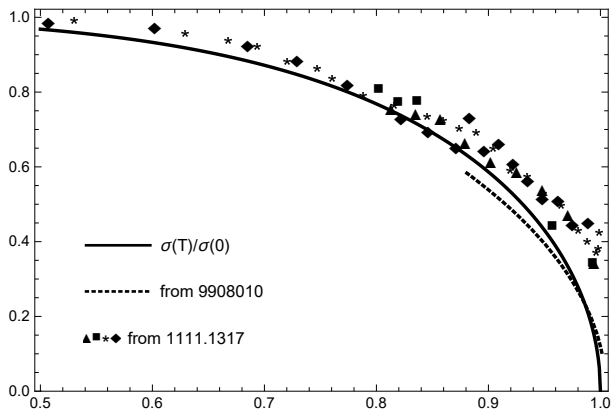}
\caption{Comparison of the lattice data for the ratio $\sigma(T)/\sigma(0)$ from \cite{30} -- dotted line, with our result from eq.(\ref{25}) -- solid line, and lattice data from \cite{30*} -- dots.}
\end{center}
\label{fig02}
\end{figure}


\vspace{1cm}

\begin{figure}[!htb]
\begin{center}
\includegraphics[width=80mm,keepaspectratio=true]{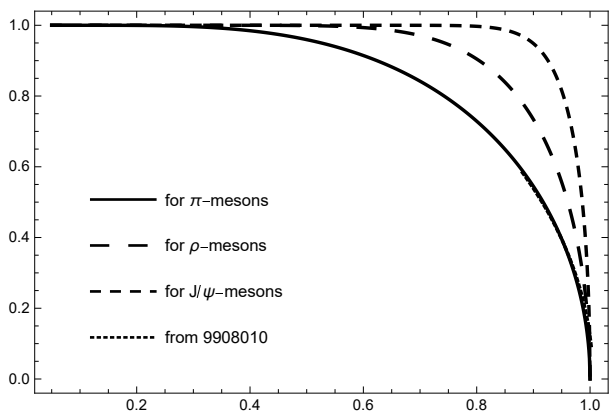}
\caption{The ratio sigma(T)/sigma(0) according to eqs.(\ref{23}),(\ref{24}) for different values of the hadron mass -- $m_{\pi}$ --solid line, $m_{\rho}$ -- long dashed line, $m_J/\psi$ -- dashed line and the same ratio from lattice data of \cite{30}-dotted line fully covered by the pion curve.}
\end{center}
\label{fig03}
\end{figure}

To understand the hadron mass dependence of different hadron curves in the Fig 2  one obtains from \ref{23}  that
$(\sigma(T)/\sigma(0))^2= 1 - (T/T_c)^{5/2} \exp{-\phi(T)},\phi(T)= \frac{M(T_c-T)}{T_c}$ which shows that the interval of
nonzero $\phi(T)$ near $T=T_c$ strongly decreases with the growing mass $M$.
Another possible comparison can be done with the Nambu--Goto type approach in \cite{59} which yields
$ \sigma(T)= \sigma(0)\sqrt{1- (T/T_c)^2}$ however disagree with the lattice data. There is a good agreement of eq.(\ref{23}) for $T \approx T_c$, which can be seen in Fig. (\ref{fig02}) comparing with the lattice data of \cite{30}.
To conclude this section we can describe the behavior of the gluonic condensate $G_2(T)$   below $T_c$ using its connection  with $\sigma_E(T)$ given in eq. (\ref{04}) yielding the approximate equation (neglecting the $T$ dependence of $ \langle \alpha_s(T) \rangle$

\be
G_2(T)=G_2(0) (1 - P_h(T)/P_h(T_c))  \label{26} \ee

where $G_2(T)$ is the colorelectric part of the total gluonic condensate, while the colormagnetic part of $G_2(T)$ is growing  together with the CM string tension as shown in the FCM \cite{40} and on the lattice \cite{7,8,9}.

\section{The temperature dependence of the quark condensate below $T_c$}

The behavior of the quark condensate $\Sigma(T)$ as a function of temperature $T$ was a topic of intensive investigations for the last
decades since it provided information on the important issue of the chiral symmetry breaking and the chiral physics in general. It was found that $\Sigma(T)$ behaves in general similarly to the CE string tension as a function of
$T$ for small quark masses $m_q$ but absolute values of $\Sigma(m_q,T)$ strongly decrease for large $m_q$. In the framework of the FCM the two phenomena: the quark confinement and the quark condensate - can be directly connected since the quark condensate as the quark Green's function $S(x,x)$ at one point is considered as a closed circular-like quark trajectory with the confining surface inside.This allows to write the condensate via the quadratic Green's function $G(x,y)$ and its mass eigenvalues $M_q$ as $\Sigma_q=N_c (m_q +M_q)G(x,x)$ and expanding $G(x,x)$ in the infinite set of eigenfunctions $\phi_n$ one obtains as in \cite{61}
\be
 \Sigma_q= N_c \lan m_q+ M_q\ran  \sum_n \frac{|\phi_n(0)|^2}{M_n}, \label{27}\ee

where $\phi_n(r)$ and $M_n$ are the pseudoscalar (PS) $q\bar q$ eigenfunctions and eigenvalues of the QCD Hamiltonian.
The latter are expressed via the string tension and $\alpha_s$ and neglecting the latter in the first approximation one
obtains a simple connection

\be
\Sigma_q(T)= c (\sigma_E(T))^{3/2} , \label{28} \ee

This relation can be extended to the case of nonzero magnetic field \cite{61}.
Coming back to the case without magnetic field,
one can find the ratio $\frac{\sigma(T)}{\sigma(0)}= \eta(T)$ using (\ref{25})  which gives $\eta(T)=(1-(T/T_c)^4)^{1/2}$
and finally one obtains the deconfining behavior of the quark condensate
\be
 K_q^{th}(T)=\Sigma_q(T)/\Sigma_q(0)= \left(\frac{\sigma(T)}{\sigma(0)}\right)^{3/2}= (1- (T/T_c)^4)^{3/4} .\label{29} \ee

\begin{figure}[!htb]
\begin{center}
\includegraphics[width=80mm,keepaspectratio=true]{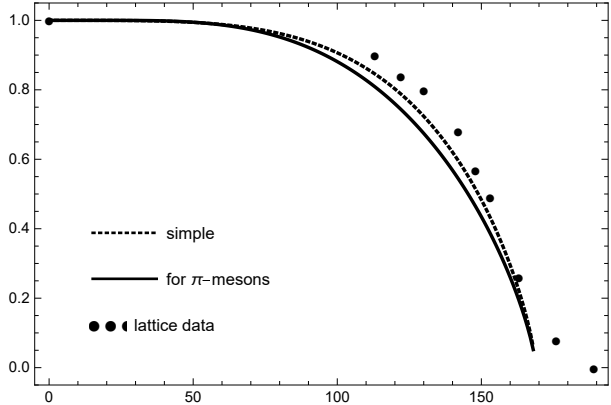}
\caption{The behavior of the quark condensate as a function of temperature T:the dotted line -- the simple form of eq.(\ref{29}),
the solid line corresponds to the string tension in the pionic hadron gas to the power $4/3$, dots-lattice data from \cite{66}.}
\end{center}
\label{fig04}
\end{figure}

 The resulting values of $K_q^{th}$ and the corresponding lattice values $K_q^{lat}$  from \cite{66} are given below in Table 3 and the whole dependence is shown in the Fig. (\ref{fig04}).

\begin{table}[!htb]
\caption{The temperature dependence of the quark condensate ratio $K_q(T)=\frac{\Sigma(T)}{\Sigma(0)}$ from eq.(\ref{29}) in comparison with the lattice data from \cite{66}}
\begin{center}
\label{tab03}
\begin{tabular}{|l|c|c|c|c|c|c|c|c|c|c|}
\hline
$T$(in MeV) & 0 &113&122&130&142&148&153&163&176&189\\
$K_q^{lat}(T,0)$  & 1 &0.90&0.84&0.80&0.68&0.57&0.49&0.26&
0.08&0\\
$K_q^{th}(T,0)$ &1 & 0.85& 0.79& 0.72& 0.6& 0.51& 0.43& 0.22& 0& 0\\

\hline
\end{tabular}
\end{center}
\end{table}


A similar behavior can be seen in the Fig. 1 of \cite{67} where the quark condensate $\Sigma = \operatorname{const} \langle \bar{q} q \rangle$ is found on the lattice in the region $T= 135-175$ MeV. The resulting values of $\Sigma_{lat}(T)$ and $\Sigma_{th}(T)$ are given below in the Table 4 where $\Sigma_{th}(T)= \Sigma_{th}(0) (1- (T/T_c)^4)^{3/4}$, $\Sigma_{th}(0)= 30$, and $T_c= 170$ MeV which is in agreement with our prediction in eq.(\ref{21}) and the lattice data in \cite{54,55,56,57,58}.

\begin{table}[!htb]
\caption{The average light quark condensate as a function of temperature T in our equation (\ref{29}) vs lattice data from \cite{67}.}
\begin{center}
\label{tab04}
\begin{tabular}{|l|c|c|c|c|c|}
\hline
$T(MeV)$ &135&145&155&165&175\\
$\Sigma_{th}$&20.5&17&12.4&5.8&0\\
$\Sigma_{lat}$&22&18&13&8&5\\
\hline
\end{tabular}
\end{center}
\end{table}
One can see in the Table 4 a reasonable agreement between the lattice data of \cite{67} and our prediction in \ref{28} supporting the strong connection between the mechanisms of confinement and the chiral symmetry breaking.

\section{Discussion of the results and conclusions}\phantom{}

(1) The main idea of the present paper is to formulate the basic element of the QCD dynamics which defines its main properties: confinement, chiral symmetry breaking (CSB), and their development with the growing temperature -- i.e. $\sigma(T)$, $\langle \bar{q}{q} \rangle (T)$, $T_c$. As it was shown above in the paper this basic element can be associated with the gluon vacuum condensate $G_2(T)$ ,its important role was already demonstrated in \cite{49,49*} and the direct connection of $G_2(T)$ and $\sigma(T)$ was found in \cite{53}. The simple mechanism VDM suggested in this paper unifies in one equation (\ref{03}) both the basic QCD scale $G_2(T)$ and the hadron pressure $P_h(T)$ and in this way $\sigma(T)$, $\langle \bar{q}{q} \rangle (T)$, $T_c$ can be found from the hadron pressure (theoretical, experimental or lattice data).

The present paper provides the main results of the proposed Vacuum Dominance Mechanism (VDM)scenario in the behavior of the gluon condensate, quark condensate and the colorelectric (CE) string tension as functions of temperature below the deconfinement transition. The main part of this connection is the "equilibrium condition" between the vacuum condensates and the pressure in eqs.(\ref{01}), (\ref{02}), (\ref{03}) which ensures confinement for $T<T_c$ while the pressure grows with $T$ at the expense of the declining vacuum store, so that the sum of the CE vacuum gluon condensate and the hadron pressure is kept constant. The exact connection of the gluon condensate and the string tension found in \cite{53} and discussed in Section 2 makes the whole picture of confinement internally connected with the basic scale and the phenomenon of the vacuum energy condensate $G_2$ introduced in \cite{49,49*,50,51,52}.

(2) The direct check of this connection has been done during the last 30 years in the phenomenological calculations of the deconfining temperature $T_c$ in \cite{40,40*,40**,41,42} which perfectly agreed with the corresponding lattice data for all studied hadron systems and also with the imposed magnetic field. In the present paper this mysterious agreement is explained within the rigorously formulated dynamical scheme called VDM which however is not yet derived from the basic principles but is imposed as an additional mechanism. Another important point of this paper is the connection of the gluon condensate and the string tension, which was previously established in \cite{28} and presented and discussed in the Section 2. The dynamical picture of the colorelectric (CE) confinement can be understood from the basic equation for the CE field correlator $D_E(x,y$ the integral of which defines string tension, and the basic interaction inside it is ensured again by confinement \cite{53}. As a result the confining factors compensate each other at distances beyond $0.2 Fm$ and the basic role in the equation for the string tension $\sigma_E$ is played by the small distance
dynamics and the gluon condensate $G_2(T)$. In Fig. 1 is shown the behavior of the $D_E(x-y)$ at large distances (beyond the maximum) which is fully defined by the small distance parameters before the maximum point.

Here the main point developed in our paper is the $T$ dependence of the gluon condensate $G_E(T)$ which imposes the corresponding $T$ dependence of the colorelectric string tension $\sigma_E(T)$ shown in eq.(\ref{07}). As a consequence of the VDM the gluon condensate decreases together with $\sigma_E(T)$ for the growing $T$  and one obtains the average behavior shown in eq.(\ref{25}) $\sigma^2_E(T)=\sigma^2_E(0)(1-(T/T_c)^4$  which is compared in Fig. 2 with the lattice data of \cite{30} showing a good agreement. A similar type of behavior  one can see in the Fig.3 where the same data were compared with the string tension in different hadronic systems, including the pions. This typical fast decreasing asymptotics of $\sigma_E(T)$ for $T$ approaching $T_c$ is presented in eqs.(\ref{24}),(\ref{25}) which explain the numerical lattice data. This result was derived to our knowledge for the first time in the literature and can be easily extended to the cases of more general thermodynamical ensembles including magnetic and electric fields which can be important for astrophysics.

(3) In Section 5 we have discussed another important aspect of the deconfinement phenomena-the disappearance of the quark (chiral) condensate $\langle \bar{q}{q} \rangle (T)$ with the growing temperature $T$ .This topic has a long story and numerous theoretical, experimental and lattice studies, e.g. the high level of the latter can be found in \cite{62,63,64,65,66,67}. One of the main topics in this field is the temperature $T_q$ of the quark condensate vanishing (the chiral transition point) and the the form of its vanishing near $T_c$. This topic was studied in the framework of the nonperturbative QCD and the quark confinement in \cite{61} (see also the cited there papers) and it was found that the chiral and confinement phenomena are closely related and in the first approximation they disappear at the same temperature, $T_q= T_c$. Moreover, the forms of the behavior of the chiral condensate and string tension are dimensionally connected:from eq.(\ref{28}) one has $\frac{\langle \bar q q \rangle(T)}{\langle \bar{q}{q} \rangle(0)}= (\frac{\sigma_E(T)}{\sigma_E(0)})^{3/2}$. This behavior is checked in Fig. (\ref{fig04}), where one can see a reasonable agreement of this law with the lattice data of \cite{66} (except for a narrow region near $T_c$ where additional effects are present). The same type of agreement can be seen in the Table 4.

As a result one can note that the close dynamical connection of the two main phenomena -- the confinement and the chiral symmetry breaking has been again demonstrated in the paper, see also a recent paper on the underlying dynamics of these phenomena in \cite{68} and the references therein.
  
(4) Summarizing the main points of the paper one can stress the significance of the proposed Vacuum Dominance Mechanism (VDM) which allows to connect the growing hadron pressure in any hadron system with the decreasing gluonic and quark condensates and in this way to solve the problem of the colorelectric deconfinement and the chiral symmetry restoration. There are many possible developments and applications of this approach involving the inclusion of baryon density, external magnetic field etc which are planned for the future. At the same time one needs the rigorous derivation of the VDM  from the fundamental field theory equations at finite temperature.

The authors are grateful to A. M. Badalian for useful criticism and to N.P. Igumnova for collaboration.

\end{document}